\documentclass[preprint2]{exoplanet}
\usepackage{times}

\newcommand{\refs}{\par\noindent\hangindent=1pc\hangafter=1}
\def\d{{\rm d}}       

\def\i{\ifmmode{\rm i}\else\char"10\fi} 
\def\frac#1#2{{#1\over#2}}              

\begin{document}

\title{\textbf{\LARGE Keplerian Orbits and Dynamics of Exoplanets}}

\author {\textbf{\large Carl D. Murray}}
\affil{\small\em Queen Mary University of London}

\author {\textbf{\large Alexandre C.M. Correia}}
\affil{\small\em University of Aveiro}

\begin{abstract}
\begin{list}{ } {\rightmargin 1in}
\baselineskip = 11pt
\parindent=1pc
{\small Understanding the consequences of the gravitational interaction between
a star and a planet is fundamental to the study of exoplanets.  The solution of
the two-body problem shows that the planet moves in an elliptical path around
the star and that each body moves in an ellipse about the common center of mass.
 The basic properties of such a system are derived from first principles and
 described in the context of detecting exoplanets. 
 \\~\\~\\~}
 
\end{list}
\end{abstract}

\section{INTRODUCTION}

The motion of a planet around a star can be understood in the context of the
two-body problem, where two bodies exert a mutual gravitational effect on each
other.  The solution to the problem was first presented by Isaac Newton (1687)
in his {\it Principia}.  He was able to show that the observed elliptical path
of a planet and the empirical laws of planetary motion derived by Kepler (1609,
1619) were a natural consequence of an inverse square law of force acting
between a planet and the Sun.  According to Newton's universal law of
gravitation, the magnitude of the force between any two masses $m_1$ and $m_2$
separated by a distance $r$ is given by 
\begin{eqnarray}
F=G{{m_1m_2}\over{r^2}}
\label{eq01}
\end{eqnarray}
where $G=6.67260\times10^{-11} {\rm Nm}^2{\rm kg}^{-2}$ is the {\it universal
gravitational constant\/}.  The law is applicable in a wide variety of circumstances. 
For example, the two bodies could be a moon orbiting a planet or a planet orbiting a
star.  Newton's achievement was to show that motion in an ellipse is the natural
consequence of such a law.  A more difficult task is to find the position and velocity of
an object in the two-body problem; this is commonly referred to as {\it Kepler problem}. 
In this chapter we derive the basic equations of the two-body problem and solve them to
show how elliptical motion arises.  We then proceed to solve the Kepler problem showing
how motion around the common center of mass of the two-body system can be used to infer
the presence of planetary companions to a star. Finally we give a few representative
examples among extra-solar planets already detected. For the most part we follow the approach of Murray \& Dermott (1999).

\bigskip
\section{BASIC EQUATIONS}

\begin{figure}[t]
\epsscale{1.}
\plotone{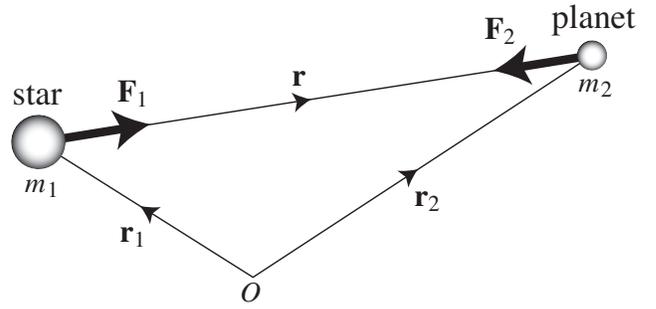}
\caption{\small The forces acting on a star of mass $m_1$ and a planet of mass $m_2$ with position vectors ${\bf r}_1$ and ${\bf r}_2$.}
\label{fig01}
\end{figure}

Consider a star and a planet of mass $m_1$ and $m_2$, respectively, with position vectors ${\bf
r}_1$ and ${\bf r}_2$ referred to an origin $O$ fixed in inertial space
(Fig.~\ref{fig01}).  

\begin{figure}[t]
\epsscale{1.}
\plotone{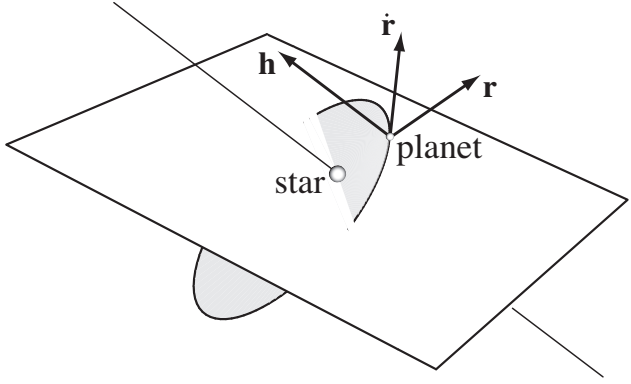}
\caption{\small The motion of $m_2$ with respect to $m_1$ defines an orbital plane (shaded region), because ${\bf r}\times\dot{\bf r}$ is a constant vector, ${\bf h}$, the angular momentum vector and this is always perpendicular to the orbit plane.}
\label{fig03}
\end{figure}

The relative motion of the planet with respect to the star is given by the vector ${\bf r}={\bf r}_2-{\bf r}_1$.  The gravitational
forces acting on the star and the planet are
\begin{eqnarray}
\label{eq02a}
{\bf F}_1  =m_1\ddot{\bf r}_1 =  +G{{m_1m_2}\over{r^3}}{\bf r}\, ,\\
\label{eq02b}
{\bf F}_2  =m_2\ddot{\bf r}_2=  -G{{m_1m_2}\over{r^3}}{\bf r}
\end{eqnarray}
respectively.  
Now consider the motion of the planet $m_2$ with respect to the star $m_1$.   If
we write $\ddot{\bf r}=\ddot{\bf r}_2-\ddot{\bf r}_1$ we can use Eq.~(\ref{eq01}) to
obtain 
\begin{eqnarray}
\ddot{\bf r}+G(m_1+m_2){{\bf r}\over{r^3}}=0\,.
\label{eq03}
\end{eqnarray}
If we take the vector product of ${\bf r}$ with Eq.~(\ref{eq03}) we have ${\bf
r}\times\ddot {\bf r}=0$ which can be integrated directly to give
\begin{eqnarray}
{\bf r}\times\dot{\bf r}={\bf h}
\label{eq04}
\end{eqnarray}
where ${\bf h}$ is a constant vector which is simultaneously perpendicular to both ${\bf r}$ and
$\dot{\bf r}$.  Therefore the motion of the planet about the star lies in a plane (the {\it orbit plane}) 
perpendicular to the direction defined by ${\bf h}$.  Another consequence of this result is
that the position and velocity vectors will always lie in the same plane (see
Fig.~\ref{fig03}).  Equation (\ref{eq04}) is often referred to as the {\it angular momentum
integral\/} and ${\bf h}$ represents a constant of the two-body motion. 

In order to solve Eq.~(\ref{eq03}) we transform to a polar coordinate system $(r,\,\theta)$ referred to an origin
centered on the star with an arbitrary reference line corresponding to
$\theta=0$.  In polar coordinates the position, velocity and acceleration vectors can be written as  
\begin{eqnarray}
\label{eq05a}
{\bf r}&=&r \, \hat{\bf r}\\
\label{eq05b}
\dot{\bf r}&=&\dot r \, \hat{\bf r}+r \dot\theta \, {\hat\theta}\\
\label{eq05c}
\ddot{\bf r}&=&(\ddot r-r\dot\theta^2)\hat{\bf r}
+\left[{{1}\over{r}}{{\d}\over{\d t}}\left(r^2\dot\theta\right)\right]
{\hat\theta}\,.
\end{eqnarray}
where $\hat{\bf r}$ and $\bf{\hat\theta}$ denote unit
vectors along and perpendicular to the radius vector respectively.
Substituting Eq.~(\ref{eq05b}) into Eq.~(\ref{eq04}) gives 
${\bf h}=r^2\dot\theta \, \hat{\bf z}$, where $\hat{\bf z}$ is a unit vector
perpendicular to the plane of the orbit forming a right-handed triad with
$\hat{\bf r}$ and $\hat\theta$.  The magnitude of this vector gives us 
\begin{eqnarray}
h=r^2\dot\theta\,.
\label{eq06}
\end{eqnarray}
Therefore, although $r$ and $\theta$ vary as the planet moves around the star, the quantity $r^2\dot\theta$ remains constant. 
The area element $d A$ swept out by the star-planet radius
vector in the time interval $d t$ is given in polar coordinate by 
\begin{eqnarray}
d A = \int_0^r r \, d r \, d \theta = {{1}\over{2}}r^2 d \theta \, ,
\label{eq07}
\end{eqnarray}
and thus
\begin{eqnarray}
\dot A={{1}\over{2}}r^2\dot\theta={{1}\over{2}}h=\hbox{ constant}\,.
\label{eq08}
\end{eqnarray} 
This is equivalent to Kepler's second law of planetary motion which states that the star-planet line sweeps out equal areas in
equal times.

Using Eq.~(\ref{eq05a}) and comparing the $\hat{\bf r}$ components of
Eqs.~(\ref{eq03}) and (\ref{eq05c}) gives the scalar differential equation
\begin{eqnarray}
\ddot r-r\dot\theta^2=-{{G(m_1+m_2)}\over{r^2}}\,.
\label{eq09}
\end{eqnarray}

\begin{figure}[t]
\epsscale{1.}
\plotone{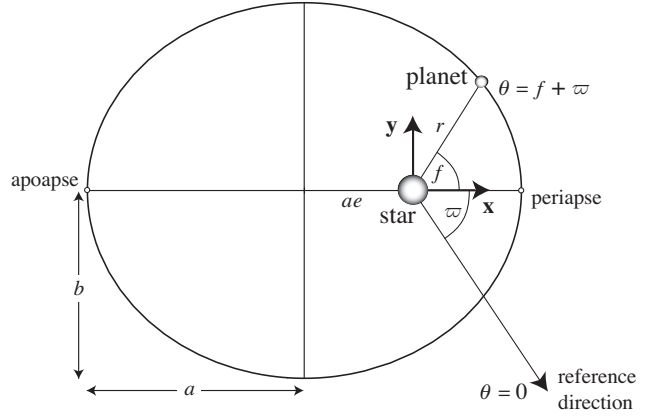}
\caption{\small The geometry of the ellipse of semi-major axis $a$, semi-minor axis $b$, eccentricity $e$ and longitude of periapse
$\varpi$.}
\label{fig04}
\end{figure}

In order to find $r$ as a function of $\theta$ we need to
make the substitution $u=1/r$.  By differentiating $r$ with respect to
time and making use of Eq.~(\ref{eq06}) we can eliminate time in the differential equation.  We obtain
\begin{eqnarray}
\ddot r =-h{{\d^2u}\over{\d\theta^2}}\dot\theta
=-h^2u^2{{\d^2u}\over{\d\theta^2}}
\label{eq10}
\end{eqnarray}
and hence Eq.~(\ref{eq09}) can be written
\begin{eqnarray}
{{\d^2u}\over{\d\theta^2}}+u={{G(m_1+m_2)}\over{h^2}}\,.
\label{eq11}
\end{eqnarray}
This is a second order, linear differential equation, often referred as {\it Binet's
equation}, with a general solution
\begin{eqnarray}
u={{G(m_1+m_2)}\over{h^2}}\left[1+e\cos(\theta-\varpi)\right] \, ,
\label{eq12}
\end{eqnarray}
where $e$ (an amplitude) and $\varpi$ (a phase) are two constants of
integration.  Substituting back for $r$ gives
\begin{eqnarray}
r={{p}\over{1+e\cos(\theta-\varpi)}} \, ,
\label{eq13}
\end{eqnarray}
where $p=h^2/G(m_1+m_2)$.  This is the general equation in polar coordinates of a set of curves known as {\it conic sections} where $e$ is the {\it eccentricity\/} and $p$ is a constant called the {\it semilatus rectum\/}.  For a given system the initial conditions will determine the particular conic section (circle, ellipse, parabola, or hyperbola) the planet follows.  We consider only elliptical motion for which
\begin{eqnarray}
p=a(1-e^2) \, ,
\label{eq14}
\end{eqnarray}
where $a$, a constant, is the {\it semi-major axis\/} of the ellipse.  The quantities $a$ and
$e$ are related by
\begin{eqnarray}
b^2=a^2(1-e^2) \, ,
\label{eq15}
\end{eqnarray}
where $b$ is the {\it semi-minor axis\/} of the ellipse (see Fig.~\ref{fig04}).  

Therefore, for any given value of $\theta$ the radius is calculated using the equation
\begin{eqnarray}
r={{a(1-e^2)}\over{1+e\cos(\theta-\varpi)}}\,.
\label{eq16}
\end{eqnarray}
Hence the path of the planet around the star is an ellipse with the star at one focus; this is Kepler's first law of planetary motion.  Note that in the special case where $e=0$ (a circular orbit), $r=a$ and the angle $\varpi$ is undefined.

The angle $\theta$ is called the {\it true longitude\/}.  
Equation (\ref{eq16}) shows that the minimum and maximum values of $r$ are $a(1-e)$ (at $\theta=\varpi$) and $a(1+e)$ (at $\theta=\varpi+\pi$), respectively.  These points are referred to as the  {\it periapse\/} and the {\it apoapse\/}, respectively, although for motion around a star they can also be referred to as the {\it periastron\/} and {\it apastron}.
 
The angle $\varpi$ (pronounced ``curly pi'') is called the {\it
longitude of periapse\/} or {\it longitude of periastron\/} of the planet's
orbit and gives the angular location of the closest approach with respect to the
reference direction.   If we define the {\it true anomaly\/} to be the angle
$f=\theta-\varpi$ (see Fig.~\ref{fig04}) then $f$ is measured with respect to the
periapse direction and Eq.~(\ref{eq16}) can be written
\begin{eqnarray}
r={{a(1-e^2)}\over{1+e\cos f}}\,.
\label{eq17}
\end{eqnarray}
In this case if we define a cartesian coordinate system centered on the
star with the $x$-axis pointing towards the periapse (see Fig.~\ref{fig04}), then
the position vector has components 
\begin{eqnarray}
x&=&r\cos f\\
y&=&r\sin f\,.
\label{eq18}
\end{eqnarray}

Although we have eliminated the time from our equation of motion, we can relate
the orbital period, $T$, to the semi-major axis, $a$.  The area of an ellipse is
$A=\pi a b$ and this is swept out by the star-planet line in a time, $T$. 
Hence,  from Eq.~(\ref{eq08}), $A=hT/2$ and so 
\begin{eqnarray}
T^2={{4\pi^2}\over{G(m_1+m_2)}}a^3\,.
\label{eq19}
\end{eqnarray}
This is Kepler's third law of planetary motion.  It implies that the period of the planet's orbit is independent of $e$ and is purely a function of the sum of the masses and $a$.  If we define the {\it mean motion\/}, $n$ of the planet's motion as
\begin{eqnarray}
n={{2\pi}\over{T}}
\label{eq20}
\end{eqnarray}
then we can write
\begin{eqnarray}
G(m_1+m_2)=n^2a^3
\label{eq21}
\end{eqnarray}
and hence
\begin{eqnarray}
h=na^2\sqrt{1-e^2}=\sqrt{G(m_1+m_2) a(1-e^2)}\,.
\label{eq22}
\end{eqnarray}

There is an additional constant of the two-body motion which is useful in calculating the velocity of the planet.  Taking the scalar
product of $\dot{\bf r}$ with Eq.~(\ref{eq03}) and using Eqs.~(\ref{eq05a}) and
(\ref{eq05b}) gives the scalar equation
\begin{eqnarray}
\dot{\bf r}\cdot\ddot{\bf r}+G(m_1+m_2){{\dot r}\over{r^2}}=0
\label{eq23}
\end{eqnarray}
which can be integrated to give
\begin{eqnarray}
{{1}\over{2}}v^2-{{G(m_1+m_2)}\over{r}}=C \, ,
\label{eq24}
\end{eqnarray}
where $v^2=\dot{\bf r}\cdot\dot{\bf r}$ is the square of the velocity
and $C$ is a constant of the motion.  Equation (\ref{eq24}) is called the
{\it vis viva integral\/}.  It shows that the orbital energy per unit mass of
the system is conserved.  

Because $\varpi$ is a constant, $\dot\theta=\dot f$ and Eq.~(\ref{eq05b}) gives
\begin{eqnarray}
v^2=\dot{\bf r}\cdot\dot{\bf r}={\dot r}^2+r^2{\dot f}^2\,.
\label{eq25}
\end{eqnarray}
By differentiating Eq.~(\ref{eq17}) we obtain
\begin{eqnarray}
\dot r={{r\,\dot f\,e\sin f}\over{1+e\cos f}}
\label{eq26}
\end{eqnarray}
and hence, using Eqs.~(\ref{eq06}) and (\ref{eq13}), we have 
\begin{eqnarray}
\dot r={{na}\over{\sqrt{1-e^2}}}\,e\sin f
\label{eq27}
\end{eqnarray}
and
\begin{eqnarray}
r\dot f={{na}\over{\sqrt{1-e^2}}}(1+e\cos f)\,.
\label{eq28}
\end{eqnarray}
Therefore we can write Eq.~(\ref{eq25}) as
\begin{eqnarray}
v^2={{n^2a^2}\over{1-e^2}}(1+2\,e\cos f+e^2)\,.
\label{eq29}
\end{eqnarray}
This shows the dependence of $v$ on $f$.  A little further manipulation gives
\begin{eqnarray}
v^2=G(m_1+m_2)\left({{2}\over{r}}-{{1}\over{a}}\right)
\label{eq30}
\end{eqnarray}
which shows the dependence of $v$ on $r$.

\bigskip
\section{SOLUTION OF THE KEPLER PROBLEM}

In the previous section we solved the equation of motion of the two-body problem
to show the path of the planet's orbit with respect to the star.  However, in
the process we eliminated the time and so although we can calculate $r$ for a
given value of $\theta$, we have no means of finding $r$ as a function of time.
This is the essence of the {\it Kepler problem}. 

Our starting point is to derive an expression for $\dot r$ in terms of $r$.  We
can do this by using Eqs.~(\ref{eq17}), (\ref{eq28}) and (\ref{eq30}) to rewrite Eq.~(\ref{eq25}) as
\begin{eqnarray}
{\dot r}^2=n^2a^3\left({{2}\over{r}}-{{1}\over{a}}\right)
-{{n^2a^4(1-e^2)}\over{r^2}}\,.
\label{eq31}
\end{eqnarray}
This simplifies to give
\begin{eqnarray}
\dot r={{na}\over{r}}\sqrt{a^2e^2-(r-a)^2}\,.
\label{eq32}
\end{eqnarray}

In order to solve this differential equation we introduce a new variable, $E$, the {\it eccentric anomaly,\/} by means of the substitution
\begin{eqnarray}
r=a(1-e\cos E)\,.
\label{eq33}
\end{eqnarray}
The differential equation transforms to
\begin{eqnarray}
\dot E={{n}\over{1-e\cos E}}\,.
\label{eq34}
\end{eqnarray}
The solution can be written as
\begin{eqnarray}
n(t-t_0)=E-e\sin E \, ,
\label{eq35}
\end{eqnarray}
where we have taken $t_0$ to be the constant of integration and used
the boundary condition $E=0$ when $t=t_0$.  At this point we can define a new quantity, $M$, the {\it mean anomaly\/} such that
\begin{eqnarray}
M=n(t-t_0) \, ,
\label{eq36}
\end{eqnarray}
where ${t_0}$ is a constant called the {\it time of periastron passage\/}. 
There is no simple geometrical interpretation of $M$ but we note that it has the
dimensions of an angle and that it increases linearly with time.  Furthermore,
$M=f=0$ when $t=t_0$ or $t=t_0+T$ (periapse passage) and $M=f=\pi$ when
$t=t_0+T/2$ (apoapse passage).  We can write 
\begin{eqnarray}
M=E-e\sin E\,.
\label{eq37}
\end{eqnarray}
This is {\it Kepler's equation\/} and its solution is fundamental to the problem
of finding the orbital position at a given time.  For a particular time $t$ we
can (i) find $M$ from Eq.~(\ref{eq36}), (ii) find $E$ by solving Kepler's equation,
Eq.~(\ref{eq37}), (iii) find $r$ using Eq.~(\ref{eq33}), and finally $f$ using
Eq.~(\ref{eq17}).   

The key step is solving Kepler's equation and this is usually done numerically. 
Danby (1988) gives several numerical methods for its solution.  For example, if
we define the function 
\begin{eqnarray}
g(E)=E-e\sin E-M
\label{eq38}
\end{eqnarray}
then we can use a Newton-Raphson method to find the root of the non-linear equation, $g(E)=0$.  The iteration scheme is
\begin{eqnarray}
E_{i+1}=E_i-{{g(E_i)}\over{g'(E_i)}},\qquad i=0,1,2,\dots
\label{eq39}
\end{eqnarray}
where $g'(E_i)=\d g(E_i)/\d E_i=1-e\cos E_i$ and the iterations proceed until
convergence is achieved. A reasonable initial value is $E_0=M$ since $E$ and $M$
differ by a quantity of order $e$ (see Eq.~(\ref{eq37})).

Although we cannot have an explicit relation between the angles $f$ and
$M$, from Kepler's second law (Eq.~\ref{eq06}) and (Eq.~\ref{eq22}) it is possible to write:
\begin{eqnarray}
\d f = n \sqrt{1-e^2} \left({{a}\over{r}}\right)^2  \d t =
\sqrt{1-e^2}  \left({{a}\over{r}}\right)^2 \d M \,.
\label{eq40}
\end{eqnarray}
The above relation is useful when we want to average any physical quantity
over a complete orbit. For instance,
\begin{eqnarray}
\left<{{1}\over{r^2}} \right> 
= {{1}\over{2 \pi}} \int_0^{2 \pi} {{\d M}\over{r^2}} = {{1}\over{a^2
\sqrt{1-e^2}}} \,. 
\label{eq41}
\end{eqnarray}

To complete the set of useful angles we define the {\it mean longitude\/}, $\lambda$ by
\begin{eqnarray}
\lambda = M+\varpi\,.
\label{eq41}
\end{eqnarray}
Therefore $\lambda$, like $M$, is a linear function of time.  It is important to note that all longitudes ($\theta$, $\varpi$, $\lambda$) are defined with respect to a common, arbitrary reference direction.

\bigskip
\section{THE ORBIT IN THREE DIMENSIONS}

Of the orbital elements we have defined so far, two ($a$ and $e$) are related to the
physical dimensions of the orbit and the remaining two ($\varpi$ and $f$) are related to
orientation of the orbit or the location of the planet in its orbit.  Note that there are
many alternatives to $f$ (e.g.~ $\theta$, $M$ and $\lambda$) and the time of
periapse passage, $t_0$, can be used instead of $f$ since the latter can always
be calculated from the former.  We have already noted that $\varpi$ is the
angular location of the periapse direction measured from a reference point on
the orbit. 

Consider the planet's position vector, 
\begin{eqnarray}
{\bf r}=(x,y,0)=x \, \hat{\bf x}+y \, \hat{\bf y} +0 \, \hat{\bf z}
\label{eq42}
\end{eqnarray}
in a three-dimensional coordinate system where the $x$-axis lies along the major (long) axis of the ellipse in the direction of periapse, the $y$-axis is perpendicular to the $x$-axis and lies in
the orbital plane, while the $z$-axis is mutually perpendicular to both the
$x$- and $y$-axes forming a right-handed triad.  By definition orbital motion is
confined to the $x$-$y$ plane.  Consider a standard coordinate system where the
direction of the reference line in the 
reference plane forms the $X$-axis.  The $Y$-axis is in the reference plane at
right-angles to the $X$-axis, while the $Z$-axis is perpendicular to both the
$X$- and $Y$-axes 
forming a right-handed triad. 

\begin{figure}[t]
\epsscale{1.}
\plotone{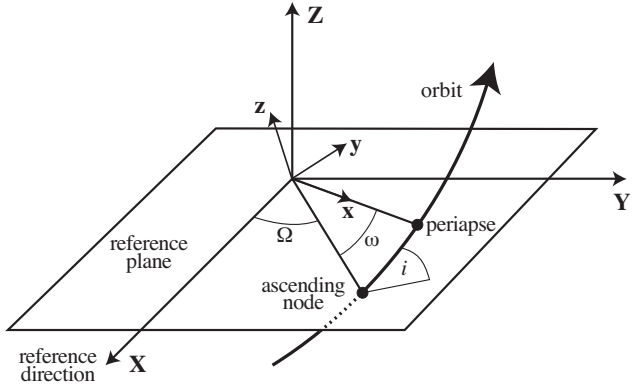}
\caption{\small The relationship between the $(x,y,z)$ and $(X,Y,Z)$ coordinate systems and the angles $\omega$, $I$ and $\Omega$. }
\label{fig05}
\end{figure}

Let $I$ denote the {\it inclination\/}, the angle between the orbit plane and
the reference plane.  The line formed by the intersection of the two planes is
called the {\it line of nodes\/}.  The {\it ascending node\/}  is the point in
both planes where the orbit crosses the reference plane moving from below to
above he plane.  The {\it longitude of ascending node\/}, $\Omega$ is the angle
between the reference line and the radius vector to the ascending node. The
angle between this same radius vector and the periapse of the orbit is called
the {\it argument of periapse\/}, $\omega$.  Note that the inclination is always
in the range $0\le I\le180^\circ$. An orbit is said to be {\it prograde\/} if
$I<90^\circ$ while if $I\ge90^\circ$ the motion is said to be {\it
retrograde\/}. We can also define   
\begin{eqnarray}
\varpi=\Omega+\omega
\label{eq43}
\end{eqnarray}
where $\varpi$ is the longitude of periapse introduced above but that now, in
general, the angles $\Omega$ and $\omega$ lie in different planes so that
$\varpi$ forms a `dog-leg' angle.   

The orientation angles $I$, $\Omega$ and $\omega$ are illustrated in Fig.~5.  
It is clear that coordinates in the $(x,y,z)$ system can be expressed in terms
of the $(X,Y,Z)$ system by means of a series of three rotations: (i) a rotation
about the $z$-axis through an angle $\omega$ so that the $x$-axis coincides with
the line of nodes, (ii) a rotation about the $x$-axis through an angle $I$ so
that the two planes are coincident and finally (iii) a rotation about the
$z$-axis through an angle $\Omega$.  
We can represent these transformations by
two $3\times3$ rotation matrices, denoted by ${\bf P}_x (\phi) $ (rotation about
the $x$-axis) and ${\bf P}_z (\phi) $ (rotation about the $z$-axis),
with elements

\begin{eqnarray}
{\bf P}_x (\phi) =\left(\matrix{
1&0&0\cr
0&\cos\phi&-\sin\phi\cr
0&\sin\phi&\cos\phi\cr}\right)
\label{eq44}
\end{eqnarray}
and

\begin{eqnarray}
{\bf P}_z (\phi) =\left(\matrix{
\cos\phi&-\sin\phi&0\cr
\sin\phi&\cos\phi&0\cr
0&0&1\cr}\right)\,.
\label{eq45}
\end{eqnarray}

Consequently

\begin{eqnarray}
\left(\matrix{X\cr Y\cr Z}\right)=
{\bf P}_z (\Omega) {\bf P}_x (I) {\bf P}_z (\omega) \left(\matrix{x\cr y\cr z}\right)
\label{eq46}
\end{eqnarray}
and

\begin{eqnarray}
\left(\matrix{x\cr y\cr z}\right)=
{\bf P}^{-1}_z (\omega) {\bf P}^{-1}_x (I) {\bf P}^{-1}_z (\Omega) \left(\matrix{X\cr Y\cr Z}\right)
\label{eq47}
\end{eqnarray}
where ${\bf P}^{-1}_x (\phi) = {\bf P}_x (- \phi)$ and ${\bf P}^{-1}_z (\phi) =
{\bf P}_z (- \phi)$ are the inverse of the matrices of 
${\bf P}_x (\phi) $ and ${\bf P}_z (\phi) $, respectively.

If we now restrict ourselves to coordinates which lie in the orbital plane, we
have $x=r\cos f$, $y=r\sin f$, $z=0$ and
\begin{eqnarray}
X&=r\left(
\cos\Omega\cos(\omega+f)-\sin\Omega\sin(\omega+f)\cos I\right)\, \\
Y&=r\left(
\sin\Omega\cos(\omega+f)+\cos\Omega\sin(\omega+f)\cos I\right)\, \\
Z&=r
\sin(\omega+f)\sin I\,.
\label{eq48}
\end{eqnarray}

\bigskip
\section{BARYCENTRIC MOTION}

\begin{figure}[t]
\epsscale{1.}
\plotone{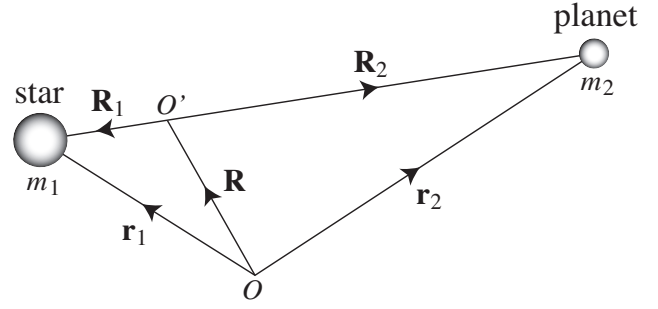}
\caption{\small The position vectors of star and planet with respect to the origin, $O$, and with respect to the center of mass of the star-planet system, $O'$.}
\label{fig02}
\end{figure}

\begin{figure}[t]
\epsscale{1.01}
\plotone{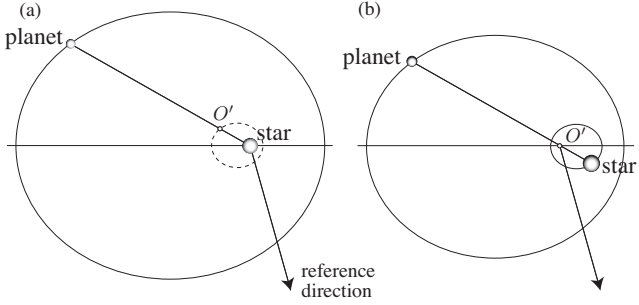}
\caption{\small (a) The motion of the planet $m_2$ with respect to
the star $m_1$ in the two-body problem; the dashed curve denotes the
elliptical path of the center of mass, $O'$. (b) The motion of the
masses $m_1$ and $m_2$ with respect to the center of mass, $O'$, for the same system.  For the purposes of illustration we used $m_2/m_1=0.2$ and $e=0.5$.}
\label{fig06}
\end{figure}

In order to determine the observable effects of an orbiting planet on a star it
helps if we consider the motion in the center of mass or {\it
barycentric system} (see Fig.\ref{fig02}).   
The position vector of the center of mass of the system is  
\begin{eqnarray}
{\bf R}={m_1{\bf r_1}+m_2{\bf r_2}\over{m_1+m_2}} \,.
\label{eq49}
\end{eqnarray}
From Eqs.~(\ref{eq02a}) and (\ref{eq02b}) we have
\begin{eqnarray}
\ddot{\bf R} = {m_1\ddot{\bf r}_1+m_2\ddot{\bf r}_2\over{m_1+m_2}} = 0 \,,
\label{eq50}
\end{eqnarray}
and by direct integration $\dot{\bf R} = {\bf V} = $~constant.
These equations imply that either (i) the center of mass is
stationary (the case when ${\bf V}=0$), or (ii) it is moving with a constant
velocity (the case when ${\bf V}\ne0$) in a straight line with respect to the
origin $O$.   
Then, if we write ${\bf R}_1={\bf r}_1-{\bf R}$ and ${\bf R}_2={\bf r}_2-{\bf
R}$, we have 
\begin{eqnarray}
m_1{\bf R}_1+m_2{\bf R}_2=0 \, .
\label{eq51}
\end{eqnarray}
This implies that (i) ${\bf R}_1$ is always in the opposite direction
to ${\bf R}_2$, and hence that (ii) the center of mass is always on the line
joining $m_1$ and $m_2$.  Therefore we can write 
\begin{eqnarray}
R_1+R_2=r \, ,
\label{eq52}
\end{eqnarray}
where $r$ is the separation of $m_1$ and $m_2$, and the distances of the star and planet
from their common center of mass are related by $m_1 R_1 = - m_2 R_2$
(Eq.~\ref{eq51}). Hence
\begin{eqnarray}
R_1={{m_2}\over{m_1+m_2}}\,r
\quad\hbox{and}\quad
R_2=-{{m_1}\over{m_1+m_2}}\,r \, .
\label{eq53}
\end{eqnarray}
Therefore each object will orbit the center of mass of the
system in an ellipse with the same eccentricity but 
the semi-major axes is reduced in scale by a factor (see Fig.~\ref{fig06})
\begin{eqnarray}
a_1={{m_2}\over{m_1+m_2}}\,a
\quad\hbox{and}\quad 
a_2={{m_1}\over{m_1+m_2}}\,a\,.
\label{eq54}
\end{eqnarray}
The orbital periods of the two objects must each be equal to $T$ and therefore
the two mean motions must also equal be equal ($n_1=n_2=n$), although the
semi-major axes are not. 
Each mass then moves on its own elliptical orbit with respect to the common
center of mass, and the periapses of their orbits differ by $\pi$ (see Fig.~6b).

\begin{figure}[t]
\epsscale{1.}
\plotone{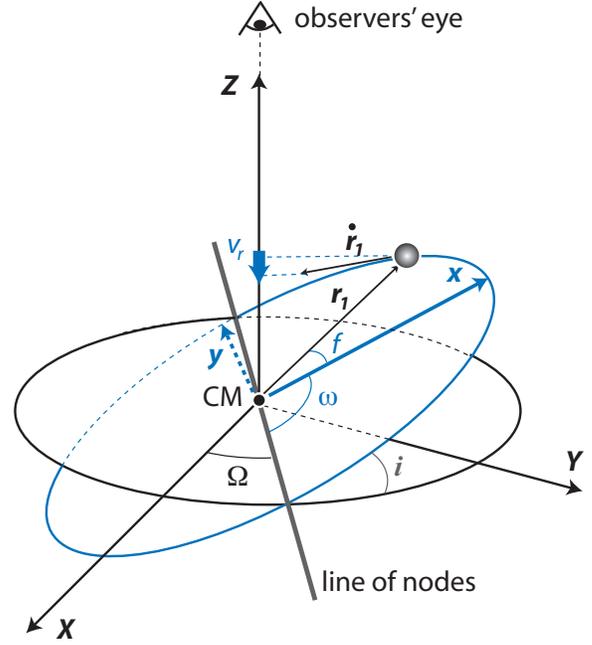}
\caption{\small The relationship between the star's velocity around the center of mass, $\dot{\bf r}_1$, and its radial component along the line of sight, $v_r$.}
\label{fig07}
\end{figure}

We are now in a position to revisit the expression for the radial velocity of
the star, $v_r$.
Observers usually take the reference plane $(X,Y)$ to be the plane of the sky
perpendicular to the line of sight, the $Z$-axis oriented towards the observer
(Fig.~\ref{fig07}).
Thus, the radial velocity of the star is simply given by the projection of the
velocity vector on the line of sight.  Since $ {\bf r_1} = {\bf R} +
{\bf R_1} $ this gives
\begin{eqnarray}
v_r = \dot{\bf r}_1 \cdot {\bf \hat{Z}} 
= V_Z + {{m_2}\over{m_1+m_2}} \dot{Z} \,, 
\label{eq55}
\end{eqnarray}
where $ V_Z = {\bf V} \cdot {\bf \hat{Z}} $ is the proper motion of the
barycenter and $\dot{Z}$ can be obtained directly from Eq.~(\ref{eq48}):
\begin{eqnarray}
\dot{Z} = \dot{r} \sin(\omega+f) \sin I + r \dot{f} \cos(\omega+f) \sin I \,, 
\label{eq56}
\end{eqnarray}
or, making use of Eqs.~(\ref{eq27}) and (\ref{eq28}),
\begin{eqnarray}
\dot Z = {n a \sin I \over{\sqrt{1-e^2}}} \left( \cos(\omega + f) + e \cos
\omega \right) \,.
\label{eq57}
\end{eqnarray}
We can now write
\begin{eqnarray}
v_r = V_Z + K \left( \cos(\omega + f) + e \cos \omega \right) \,, 
\label{eq58}
\end{eqnarray}
where
\begin{eqnarray}
K = {{m_2}\over{m_1+m_2}} {n a \sin I \over{\sqrt{1-e^2}}} \, .
\label{eq59}
\end{eqnarray}

\bigskip
\section{APPLICATION TO EXTRA-SOLAR PLANETS}

More than 500 extra-solar planets are known to date\footnote{The Extrasolar
Planets Encyclopedia. http://exoplanet.eu/}, and the number is continuously
rising.
Looking at this data, we can admire the wide variety of
possible orbital parameters and physical properties: central stars of spectral
types from F to M, minimum masses from 2 Earth-masses to more than 20 times the
mass of Jupiter, orbital periods of one day to more than fifteen
years (the same time as the length of the observations), and
eccentricities ranging from perfect circular orbits to extreme values of more
than 0.9. There are planets as close as 0.014~AU and as far as 670~AU from
their host stars.

In Table~\ref{Tab1} we report two examples of extreme values for the
eccentricity, obtained using the radial velocity technique.
The first example ({\small HD}\,156846\,b) corresponds to a highly eccentric
orbit, while the second one ({\small HD}\,83443\,b) shows an almost
circular orbit.  
At present, both planets are in single-planet systems, which allows us to
apply directly the formulae derived in previous sections to the analysis 
of their motion.

\begin{deluxetable}{l c c}
\tabletypesize{\small}
\tablecaption{Two examples of extreme orbital eccentricities for extra-solar planets.
\label{Tab1}}
\tablewidth{0pt}
\tablehead{
                        & {\small HD}\,156846\,b   & {\small HD}\,83443\,b   
}
\startdata
discovery year         & 2007			   & 2002		       \\
data ref.               & Tamuz {\it et al.} (2008)& Mayor {\it et al.} (2004) \\ \hline
\smallskip
$V_Z$ [km/s]            & $-68.540 \pm 0.001 $	   & $29.027 \pm 0.001$     \\
$T$ [d]               & $359.51 \pm 0.09 $	   & $2.98565 \pm 0.00003$     \\
$K$ [km/s]              & $ 0.464 \pm 0.003 $	   & $58.1 \pm 0.4 $  	       \\
$e$                     & $0.847 \pm 0.002 $	   & $0.013 \pm 0.013$         \\
$\omega$ [$^\circ$]          & $52.2 \pm 0.4 $	   & $ 11 \pm 11 $	       \\
$t_0$ [JD--2.45$\times 10^6$]& $3998.1 \pm 0.1 $	   & $1497.5 \pm 0.3$  	       \\ \hline
$m_1$ [$M_\odot$]       & $1.43 $		   & $0.90 $		       \\
$m_2\,\sin I$ [$M_{\rm Jup}$]& $10.45 $		   & $0.38 $		       \\
$a$ [AU]                & $0.9930 $		   & $0.03918 $  	       \\
\hline
\enddata
\smallskip
\end{deluxetable}

\subsection{{\small HD}\,156846\,b}

\begin{figure}[t]
\epsscale{1.}
\plotone{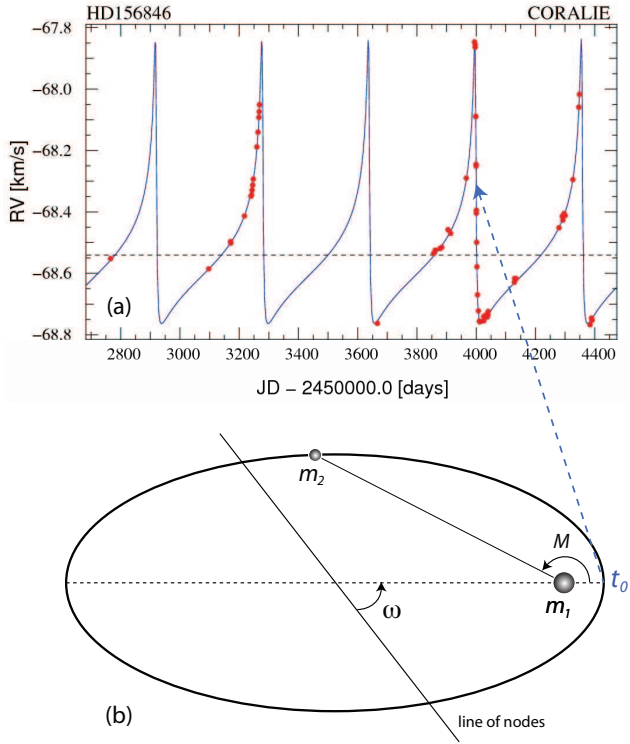}
\caption{\small (a) Radial-velocity measurements as a function of Julian
Date obtained with CORALIE for HD\,156846, superimposed on the best Keplerian
planetary solution (Table~\ref{Tab1}). (b) Keplerian orbit of HD\,156846\,b and
reference angles.}
\label{fig08}
\end{figure}

{\small HD}\,156846 has been observed with the {\small CORALIE}
spectrograph at La Silla Observatory (ESO) from May 2003 to September 2007.
Altogether, 64 radial velocity measurements with a mean uncertainty 
of 2.8~m/s were gathered.
Figure~\ref{fig08}a shows the {\small CORALIE} radial velocities and the
corresponding best-fit Keplerian model. 
The resulting orbital parameters are $T = 359.51 $~d, $e = 0.847$ and $K =
464$~m/s (Table \ref{Tab1}).
Details on the data analysis using radial velocities are given in Chapter 3.

Assuming a stellar mass $m_1 = 1.43 M_\odot$ (Tamuz {\it et al.}, 2008), Eqs.~(\ref{eq21}) and (\ref{eq59}) can be used to derive a companion minimum mass of $m_2 \sin I = 10.45 \,M_{\rm Jup}$, orbiting the central star with a semi-major axis
$a = 0.99 $~AU.
With the radial velocity technique it is impossible to determine the
inclination $I$, and therefore we are unable to describe the orbit in three
dimensions and to determine the exact mass of the planet.
Nevertheless, the orbit in two dimensions (orbital plane) can be completely
characterized.
Astrometry is the only observational technique that can provide the full, three-dimensional orbit
of the planet, but at present few planets have been observed by
this method (Chapt.~6).

In Figure~\ref{fig08}b we have drawn the orbit of {\small HD}\,156846\,b.
Because of its high eccentricity, the orbit is very elongated.
As a consequence, the separation between the planet and the star ranges
from 0.15~AU at periapse to 1.83~AU at apoapse.
In our Solar System comets are the only objects that present such large
variations in their position relative to the Sun.
The origin of such high eccentricities is unknown, but a possible
explanation is through close encounters between very massive bodies during the
formation process (Ford and Rasio, 2008).

The angle $ \omega = 52.2^\circ$ corresponds to the argument of periapse,
that is measured from the nodal line between the plane of the sky and the
orbital plane of the planet.
For {\small HD}\,156846\,b this quantity is well defined, because the orbit
is so eccentric.
According to Eq.~(\ref{eq30}), at periapse the orbital velocity is maximal and
therefore it shows an easily identifiable peak in the observational data
(Fig.~\ref{fig08}).

The planet is at periapse whenever
\begin{eqnarray}
t = t_0 + k T \quad \mathrm{with} \quad  k = 0, \pm 1, \pm 2, ... \,, 
\label{eq60}
\end{eqnarray}
where $ t_0 = \mathrm{JD}\,2453998.1 $ (19 Sep.~2006 at 14h 24m 
UT) is the time of periapse passage.
In fact, because the orbit is periodic, any instant of time, $t_0$, given by
Eq.~(\ref{eq60}) can be used as the time of periapse passage.
This is true for the two-body problem, but no longer valid if additional bodies
are present in the system.
Indeed, mutual planetary perturbations will disturb the orbits and the time of
two successive periapse passages is no longer given exactly given by Eq.~(\ref{eq60}) (see
Chapt.~10).
Although observers often use $t_0$ as a parameter to characterize orbits, for
multi-planet systems it is meaningless.
A better option is to use the mean anomaly $M$ (Eq.~\ref{eq36}) or the mean
longitude $\lambda$ (Eq.~\ref{eq41}).
The observer fixes a date $t_{f}$ and then provides the value of $M=M_0$ computed
for that date (Eq.~\ref{eq36}):
\begin{eqnarray}
M_0 = n (t_{f} - t_0) \,.
\label{eq61}
\end{eqnarray}
A practical choice of $t_f$ is to use
$t_f = t_0$, since $M_0 = 0$ (and $\lambda = \varpi $).
For multi-planetary systems, the planet will still be at periastron whenever
$M=0$, but the time of successive periapse passages will no longer be given by
Eq.~(\ref{eq60}).

\subsection{{\small HD}\,83443\,b}

\begin{figure}[t]
\epsscale{1.}
\plotone{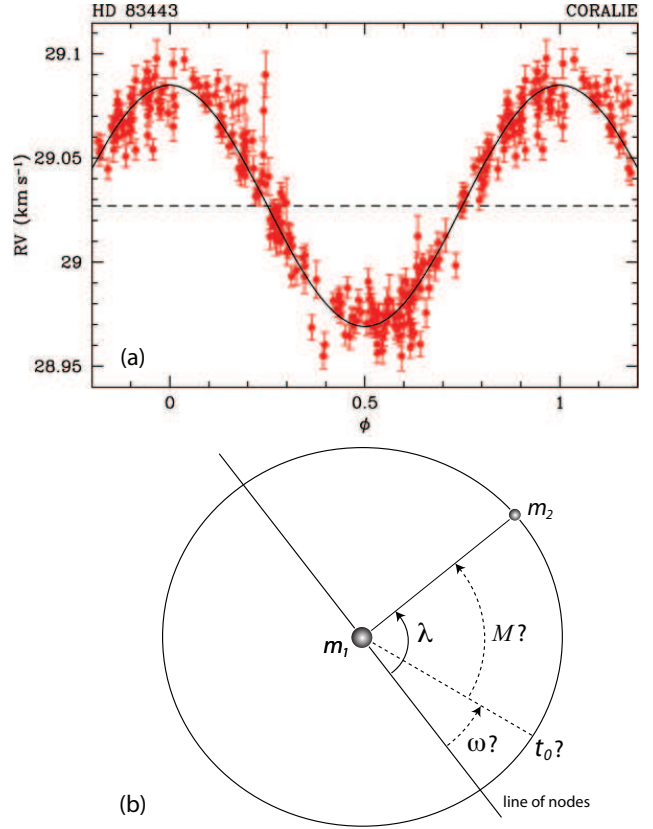}
\caption{\small (a) Phase-folded radial-velocity measurements obtained
with CORALIE for HD\,83443, superimposed on the best Keplerian
planetary solution (Table~\ref{Tab1}). (b) Keplerian orbit of HD\,83443\,b and
reference angles.}
\label{fig09}
\end{figure}

{\small HD}\,83443\,b is a short period Jupiter-size planet, therefore belonging
to the class of ``Hot-Jupiters''.
It was first announced at the Manchester IAU
Symp.~202 as a resonant 2-planet system with periods
$T_1 = 2.986 $~d and $T_2 = 29.85 $~d, but subsequent observations could not
confirm the presence of the companion around 30~d.
The origin of the transient signal is not clear yet, but an appealing
possibility is to attribute the effect to activity of the star
(Mayor {\it et al.}, 2004).
As a consequence, the {\small HD}\,83443 star has been monitored many times.
After 257 radial velocity measurements using the {\small CORALIE}
spectrograph with a mean uncertainty of 8.9~m/s (taken from March 1999 to March
2003), the short-period planet at $T = 2.986 $~d was found to revolve alone in an
almost circular orbit ($ e \sim 0 $) with $ K = 58.1 $~km/s (Table~\ref{Tab1}).
In Fig.~\ref{fig09}a we show the {\small CORALIE} radial velocities and the
corresponding best-fit Keplerian model.

Assuming a stellar mass $m_1 = 0.90 M_\odot$ (Mayor {\it et al.}, 2004),  a companion minimum mass $m_2 \sin I = 0.38 \,M_{\rm Jup}$ and semi-major
axis $ a = 0.039 $~AU has been derived.
Again, it is impossible to determine the inclination $I$, and the orbit can only
be characterized in terms of its orbital plane (Fig.~\ref{fig09}b). 

Because the orbital eccentricity is small and uncertain ($ e = 0.013 \pm
0.013 $), so too is the argument of the periapse ($ \omega = 11^\circ \pm
11^\circ $).
Indeed, for perfect circular orbits ($ e = 0 $) the argument of the periapse is
not defined since the distance from the planet to the star is constant.
Therefore, it is also meaningless to provide the time of periapse passage ($ t_0
= \mathrm{JD}\,2451497.5 \pm 0.3 $, i.e.~15 Nov.~1999 at 0h 0m UT).
The fact that the error bar is only $ 0.3 $~d, suggests (erroneously) that it is
small.
However, since the orbital period of the planet is only 2.986~d, an uncertainty of $
\pm 0.3 $~d is equivalent to a 20\% uncertainty in $ t_0 $.

For circular orbits the parameters $\omega$ and $t_0$ are not defined (because
$e=0$) and the same is also true for the mean anomaly $M$ (angle between the
periastron and the planet).
However, the orbit of the planet is still well determined (Fig.~\ref{fig09}a) and we
should be able to provide accurate positions of the planet on its orbit.
The correct parameter for that purpose is the sum $ M + \omega $ or the mean
longitude $ \lambda = M + \varpi $ (Fig.~\ref{fig09}b).
Indeed, fixing the date at $ t_f = \mathrm{JD}\,2453000.0 $ we obtain $ \lambda
= 91^\circ \pm 1^\circ $, which has a relative error of about 0.3\%, much better
than the 100\% of uncertainty in $ \omega $.

\bigskip
\bigskip
\bigskip
\textbf{Acknowledgments.} This work was partially supported by the Science
and Technology Facilities Council (UK) and by the Funda\c c\~ao para a Ci\^encia e a
Tecnologia (Portugal).

\bigskip

\bigskip
\textbf{REFERENCES}
\bigskip
\parskip=0pt
{\small
\baselineskip=11pt
\refs Danby, J.M.A.~(1988) {\em Fundamentals of Celestial Mechanics, 2nd
Edition\/}, Willmann-Bell, Richmond.

\refs Ford, E.B., and Rasio, F.A. (2008) {\em Astron.~J.\/}, {\bf
686}, 621-636.

\refs Kepler, J. (1609) {\em Astronomia Nova\/}, Heidelberg.

\refs Kepler, J. (1619) {\em Harmonices Mundi Libri V\/}, Linz.

\refs Mayor, M. {\it et al.} (2004) {\em Astron.~Astrophys.\/}, {\bf
415}, 391-402.

\refs Murray, C.D., and Dermott, S.F. (1999) {\em Solar System Dynamics\/}, Cambridge University Press, Cambridge.

\refs Newton, I. (1687). {\em Philosophiae Naturalis Principia
Mathematica\/} Royal Society, London.

\refs Tamuz, O. {\it et al.} (2008) {\em Astron.~Astrophys.\/}, {\bf
480}, L33-L36.

\end{document}